**Vortex glass scaling in Pb-doped Bi2223 single crystal**

Yu. Eltsev[a], S. Lee[b], K. Nakao[b], S. Tajima[c]

[a]P. N. Lebedev Physical Institute, RAS, Moscow, 119991, Russia

[b]Superconductivity Research Laboratory, ISTEC, 10-13, Shinonome 1-chome, Koto-ku, Tokyo, 135-0062, Japan

[c]Department of Physics, Osaka University, Toyonaka, Osaka 560-0043, Japan

**ABSTRACT**

We report on study of the vortex liquid in Pb-doped Bi-2223 single crystal using the in-plane resistivity measurements as a function of temperature and magnetic field up to 6T applied perpendicular to CuO planes. Below $T_c$ at the upper part of superconducting transition we found Arrhenius-like resistivity behavior. With further temperature decrease close to resistivity onset resistivity shows power law dependence on temperature signaling approaching vortex-glass transition. The critical exponents $\nu(z-1)=4.6\pm0.5$ are found to be field independent within experimental errors. We also present magnetic phase diagram defining region of nonzero critical current for Pb-doped Bi-2223 single crystal.

Study of high-$T_c$ superconductors behavior in a magnetic field attracts great attention due to basic scientific interests as well as need for possible practical applications. As a result of strong anisotropy and high level of thermal fluctuations mixed state magnetic phase diagram of high-$T_c$ superconductors in the rough view separated in two parts: low temperature vortex solid with a nonzero critical current density and a broad high-temperature dissipative vortex liquid phase. There are a few approaches to describe the onset of resistive transition in a magnetic field. Conventional vortex depinning process suggests that either an individual vortex line or bundles of small numbers of vortex lines are thermally activated over pinning barriers. In this single-particle picture resistivity is finite for all nonzero temperatures and follows Arrhenius behavior $R \propto \exp(-U_o/k_B T)$, where $U_o$ is the pinning barrier energy[1]. The vortex lattice melting model taking into account many-body effects [2] suggests melting transition from a low temperature ordered state into a vortex liquid. However, this model did not include effect of random pinning on melting transition. Fisher, Fisher and Huse [3] considered vortex melting transition in the presence of disorder. In the vortex-glass model second order phase transition between vortex-glass and vortex liquid occurs at a transition temperature $T_g$. Below $T_g$, in the vortex solid the vortices are frozen in a random configuration depending on kind of disorder and, consequently, are not free to move resulting in truly zero resistivity. Vortex-glass model made also several predictions concerning resistivity behavior as a function of temperature, applied magnetic field and excitation current within critical region just above $T_g$.

Vortex-glass model received substantial experimental support. In particular, the vortex-glass transition has been observed in YBCO single crystals and films [4,5], Bi-2212 single crystals [6] and other high-$T_c$ compounds. For Bi-2223 vortex solid-liquid transition has been studied in epitaxial thin films [7], c-axis oriented Bi-2223/Ag tapes [8] and Bi-2223/Bi-2212 intergrowth single crystal [9]. To get more detailed information on intrinsic properties of Bi-2223 compound in this paper we study vortex liquid properties of Pb-doped Bi-2223 single crystal probed by in-plane electrical transport measurements in dependence on temperature at several magnetic fields B//c-axis up to 6T. Our data is well fitted within vortex-glass model with field independent critical exponents $\nu(z-1)=4.6\pm0.5$. Thus, vortex-glass phase transition in Pb-doped Bi-2223 was clearly identified. Also, from our analysis we obtained phase boundary line separating vortex solid and vortex liquid states in Pb-doped Bi-2223 single crystal.

Single crystals used in this study were grown from a nominal powder composition $Bi_{1.7}Pb_{0.3}Sr_2Ca_2Cu_3O_{10+x}$ using a modified KCl flux technique at a temperature of about 850°C in magnesia crucibles [10]. After heat treatment, the KCl flux was removed by soaking crucibles in pure distilled water and the extracted crystals were further washed with dehydrated ethanol. After that single crystals were annealed at 400°C in air for 20 h to optimize $T_c$. The post-annealing was necessary because as-grown crystals were quenched or fast cooled in the furnace just after the isothermal heat treatment. A few single crystals of the typical size of 0.2x0.05x0.001 $mm^3$ grown in different batches have been chosen for measurements. Electrical contacts in four probes linear configuration were prepared by applying strips of silver paint, followed by heat treatment for about 10 min under the same conditions as during post-annealing, giving contact resistances below 0.1 Ω. To measure current-voltage response, we used usual dc-technique with excitation current 10 μA.

In Fig. 1 we present zero-field temperature dependence of the resistance of one of the samples used in this study. The inset in Fig. 1 is an enlarged view of the resistivity data near $T_c$. These data nicely illustrate high quality of our single crystals. Tc defined at a mid-point of transition is about 111K with transition width (10-90%) below 2K. Also, normal-state resistance demonstrates linear dependence on temperature and linearly extrapolated to T=0 intercepts resistance axis close to zero. These two points are often used as an empirical characterization of good sample quality and absence of c-axis resistivity contribution to the in-plane resistivity [11].

Inset in Fig. 2 shows Arrhenius plot of resistance in dependence on temperature measured at 6T. One can see substantial broadening of superconducting transition in field with resistance activated-like behavior $R \propto \exp(-U_o/k_B T)$. To analyze this behavior in more detail in the main panel of Fig. 2 we show our experimental data in the form $d\ln R/d(1/T)$ as a function of temperature. Plotted in this way our data now present temperature dependence of apparent activation energy. With decreasing temperature from 100K to about 60K apparent activation energy slightly increases and with further temperature decrease below $T^*=60K$ starts to grow much more rapidly. Similar behavior was previously observed in Bi-2212 single crystal [6] and was attributed to crossover to a critical region associated with the low-temperature three-dimensional vortex-glass phase transition.

In a critical region, according to vortex-glass model [3] resistance should follow power law dependence on temperature $R \propto (T-T_g)^{\nu(z-1)}$, where $\nu$ and $z$ are static and dynamic critical exponents correspondingly. Therefore the inverse logarithmic derivative of the resistivity $(d\ln R/dT)^{-1}$ vs temperature should be a straight line which extrapolates to zero at $T_g$ with a slope $1/\nu(z-1)$. In Fig. 3 we present data at B=6T consistent with a vortex-glass model with $T_g=35.3K$ and a slope $1/\nu(z-1)=0.22$. Power-law resistance dependence on temperature takes place inside the critical scaling region below $T^*$. $T^*$ defined in this way is approximately the same as $T^*$ extracted from the data in Fig. 2. Vortex-glass analysis for other fields used in our study gave approximately the same values of $\nu(z-1)=4.6\pm0.5$ shown in the inset in Fig. 3. Our observation is close to value $\nu(z-1)=5.2\pm0.5$ reported Bi-2223/Bi-2212 intergrowth single crystal [9]. On the other hand in previous studies of vortex-glass behavior in YBCO [4] and Bi-2212 [6] single crystal the critical exponent $\nu(z-1)=6.5\pm1.5$ has been found that is slightly higher compared to our data . Here we also note that critical scaling region observed in our study similar to Bi-2212 system [6] is rather broad, reaching about 25K at 6T. This region $\Delta T/T_g \infty 1$ is much greater compared to YBCO where more usual for critical scaling value $\Delta T/T_g \infty 0.05$ was reported [4].

Results of our I-V measurements provide further support for vortex-glass behavior in Pb-doped Bi-2223 single crystal. According to the vortex-glass model [3] the positive curvature of the I-V curves on the log-log scale indicates the vortex-liquid state, while the negative curvature reflects the vortex-glass state. Straight line with a power-law relation of logI-logV corresponds to the vortex-liquid to vortex-glass transition. In Fig.4 we present data on I-V measurements at T=60K for several magnetic fields. Similar results were observed at various temperatures in the range 40-77K. One can see change of I-V curvature from negative to positive with field increase. A dashed line separates the curves with positive and negative curvature and indicates vortex-glass transition at T=60K and field in the range 0.7-1T. Thus, results of our I-V measurements qualitatively are in a good agreement with the predictions of the vortex-glass model.

In Fig. 5 we show magnetic phase diagram constructed using results extracted from the results of R(T) and I-V measurements. Data obtained in two different ways are in a good agreement. In vortex-glass model phase transition boundary was predicted to be described as $B(T_g) \infty (T-T_g)^{2\nu_0}$, where $\nu_0=2/3$ for true 3D system. Our attempts to fit $B(T_g)$ data to this expression failed in contrast to the results obtained for c-axis oriented Bi-2223/Ag tapes and Bi-2223/Bi-2212 intergrowth single crystal in small fields below 1T [8,9].

Bi-2223 is highly anisotropic superconductor. Recently the superconducting anisotropy of Bi-2223 was reported to be about 50 [12]. Thus, Bi-2223 is not true 3D-system and some difference between our experimental data and theoretical prediction of vortex-glass model is understandable reflecting the fact of limited applicability of the vortex-glass theory to Bi-2223 system. Furthermore, picture may be much more complicated due to possible dimensional 2D-3D crossover at about 1-2T [6,7]. Our results clearly show existence of vortex liquid-solid transition in Pb-doped Bi-2223 single crystal in fields up to 6T similar to previous observations of this transition in YBCO and Bi-2212 systems [4-6]. However, understanding of true nature of this transition claims for further study.

In summary, we have studied transport properties of the vortex liquid in Pb-doped Bi-2223 single crystal by measuring current-voltage response as a function of temperature and magnetic field B//c-axis up to 6T. At the upper part of superconducting transition resistance roughly shows Arrhenius behavior. At lower temperatures apparent activation energy starts to diverge reflecting approaching vortex-glass transition. Resistance data analysis within vortex-glass model gives approximately the same critical exponents $\nu(z-1)=4.6\pm0.5$ for several fields. Results of I-V measurements also provide qualitative support for existence of vortex-glass transition. Finally, we have defined phase transition boundary separating region of nonzero critical current from dissipative vortex-liquid state for Pb-doped Bi-2223 single crystal.

This work is supported by the New Energy and Industrial Technology Development Organization (NEDO) as Collaborative Research and Development of Fundamental Technologies for Superconductivity Applications and partially supported by Russian Federal agency on science and innovation (02.513.11.3378).

**Legends to Figures:**

**Fig.1**

Zero-field temperature dependence of the resistance of Pb-doped Bi-2223 single crystal. Inset: Enlarged view of superconducting transition.

**Fig.2**

Temperature dependence of apparent activation energy at B=6T obtained as a derivative of Arrhenius plot at the same field shown in the inset. T* shows upper boundary of critical scaling region.

**Fig.3**

Main panel: (dlnρ/d(1/T) vs T at B=6T. A fit to the vortex-glass model with $T_g$=35.3K and $[\nu(z-1)]^{-1}$ is shown by the line. T* marks the onset of deviations for increasing temperatures. Inset: critical exponents ν(z-1) for different fields.

**Fig.4**

I-V curves at T=60K at several fields. Increasing field magnitudes correspond to curves from right to left.

**Fig.5**

Vortex-solid to vortex-liquid transition phase boundary extracted from resistive and current-voltage measurements. Solid line is guide for the eye.

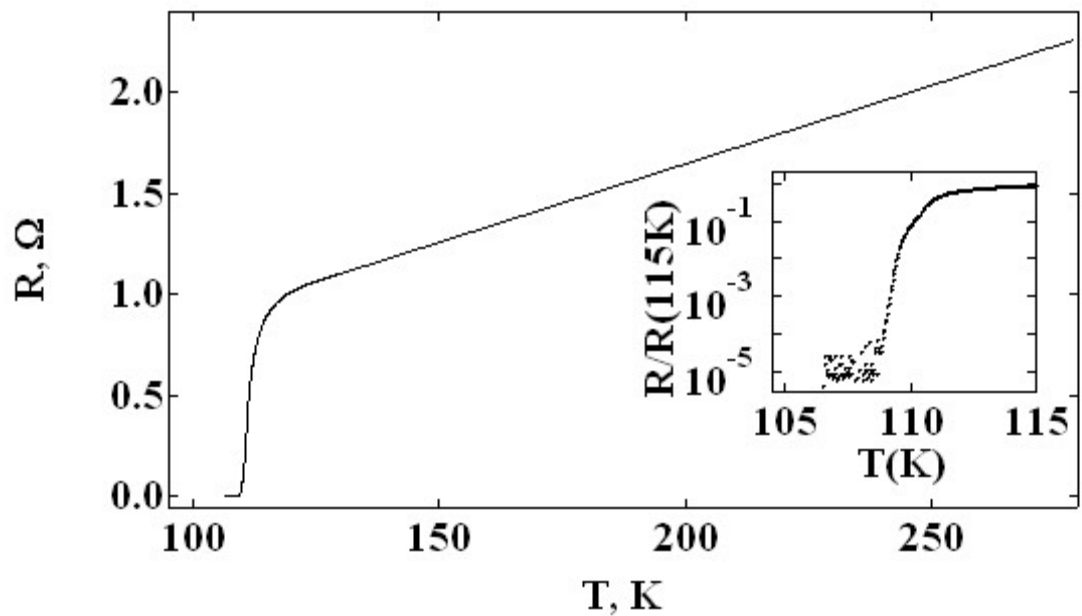

**Fig.1**

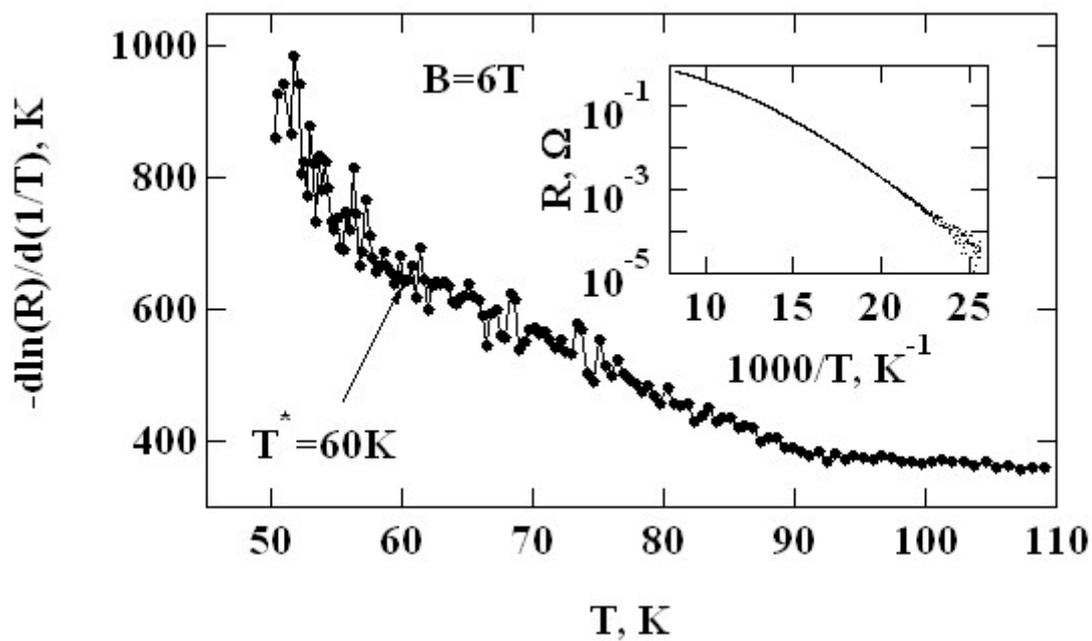

**Fig.2**

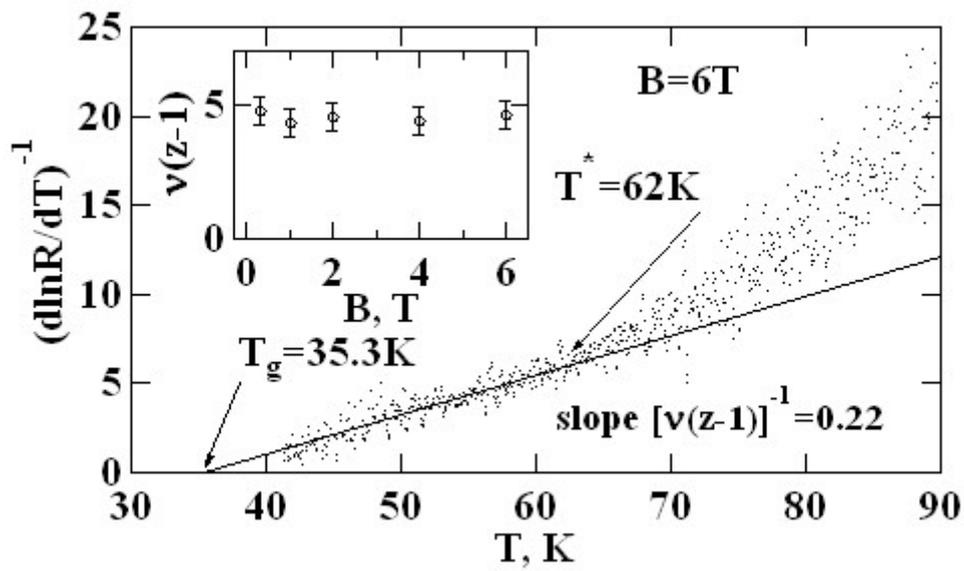

**Fig.3**

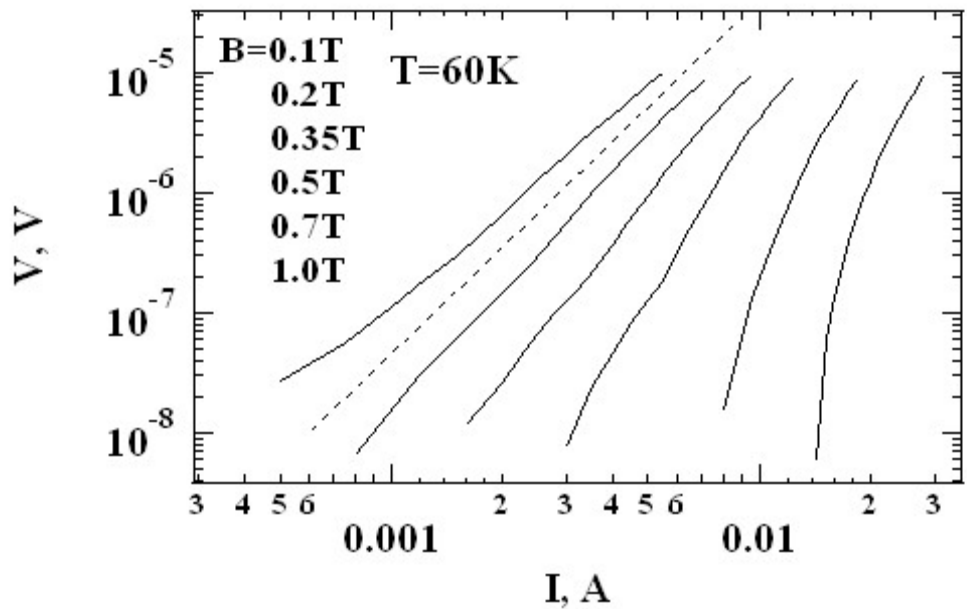

**Fig.4**

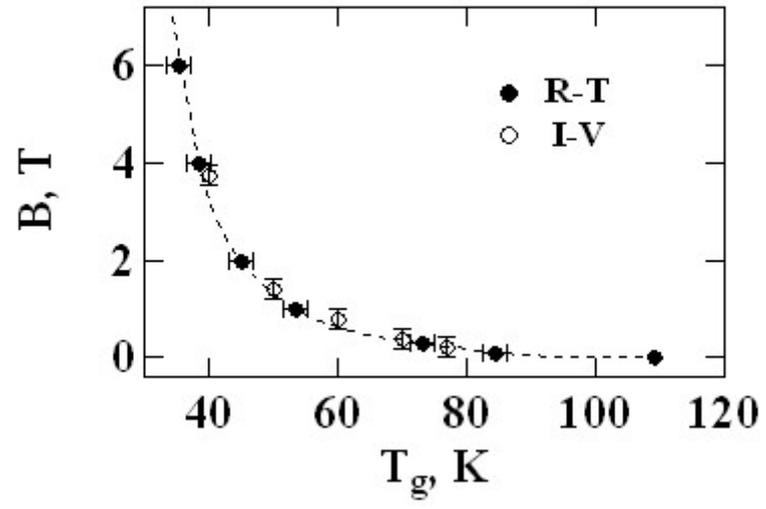

**Fig.5**